\documentclass[aps,prx,twocolumn,superscriptaddress,noeprint,longbibliography, nofootinbib,floatfix]{revtex4-2}

\usepackage{graphicx}
\usepackage{bm}
\usepackage{amsmath, amssymb}
\usepackage{booktabs}
\usepackage{mathrsfs}
\usepackage[normalem]{ulem}
\usepackage{pifont}
\usepackage[mathscr]{euscript}
\usepackage{physics}
\usepackage{xcolor}

\newcommand{\quotes}[1]{``#1''}

\begin{document}

\title{Dynamical fractal and anomalous noise in a clean magnetic crystal}

\author{Jonathan N. Hall\'en}
\address{TCM Group, Cavendish Laboratory, University of Cambridge, Cambridge CB3 0HE, UK}
\address{Max Planck Institute for the Physics of Complex Systems, 01187 Dresden, Germany}

\author{Santiago A. Grigera}
\address{Instituto de F\'{\i}sica de L\'{\i}quidos y Sistemas Biol\'ogicos, UNLP-CONICET, La Plata, Argentina}

\author{D. Alan Tennant}
	\address{Department of Physics and Astronomy, University of Tennessee, Knoxville, TN 37996, USA}
	\address{Department of Materials Science and Engineering, University of Tennessee, Knoxville, TN 37996, USA}

\author{Claudio Castelnovo}
\address{TCM Group, Cavendish Laboratory, University of Cambridge, Cambridge CB3 0HE, UK}

\author{Roderich Moessner}
\address{Max Planck Institute for the Physics of Complex Systems, 01187 Dresden, Germany}

\date{November 1, 2022}

\begin{abstract}
Fractals -- objects with non-integer dimensions -- occur in manifold settings and length scales in nature, ranging from snowflakes and lightning strikes to natural coastlines. Much effort has been expended to generate fractals for use in many-body physics.
Here, we identify an emergent dynamical fractal in a disorder-free, stoichiometric three-dimensional magnetic crystal in thermodynamic equilibrium. The phenomenon is born from constraints on the dynamics of the magnetic monopole excitations in spin ice, which restrict them to move on the fractal. This observation explains the anomalous exponent found in magnetic noise experiments in the spin ice compound Dy$_2$Ti$_2$O$_7$, and it resolves a long standing puzzle about its rapidly diverging relaxation time. 
The capacity of spin ice to exhibit such striking phenomena holds promise of further surprising discoveries in the cooperative dynamics of even simple topological many-body systems. 
\end{abstract}
\maketitle

\section{Introduction}
\begin{figure}[hb!]
    \centering
    \includegraphics[width=0.5\textwidth]{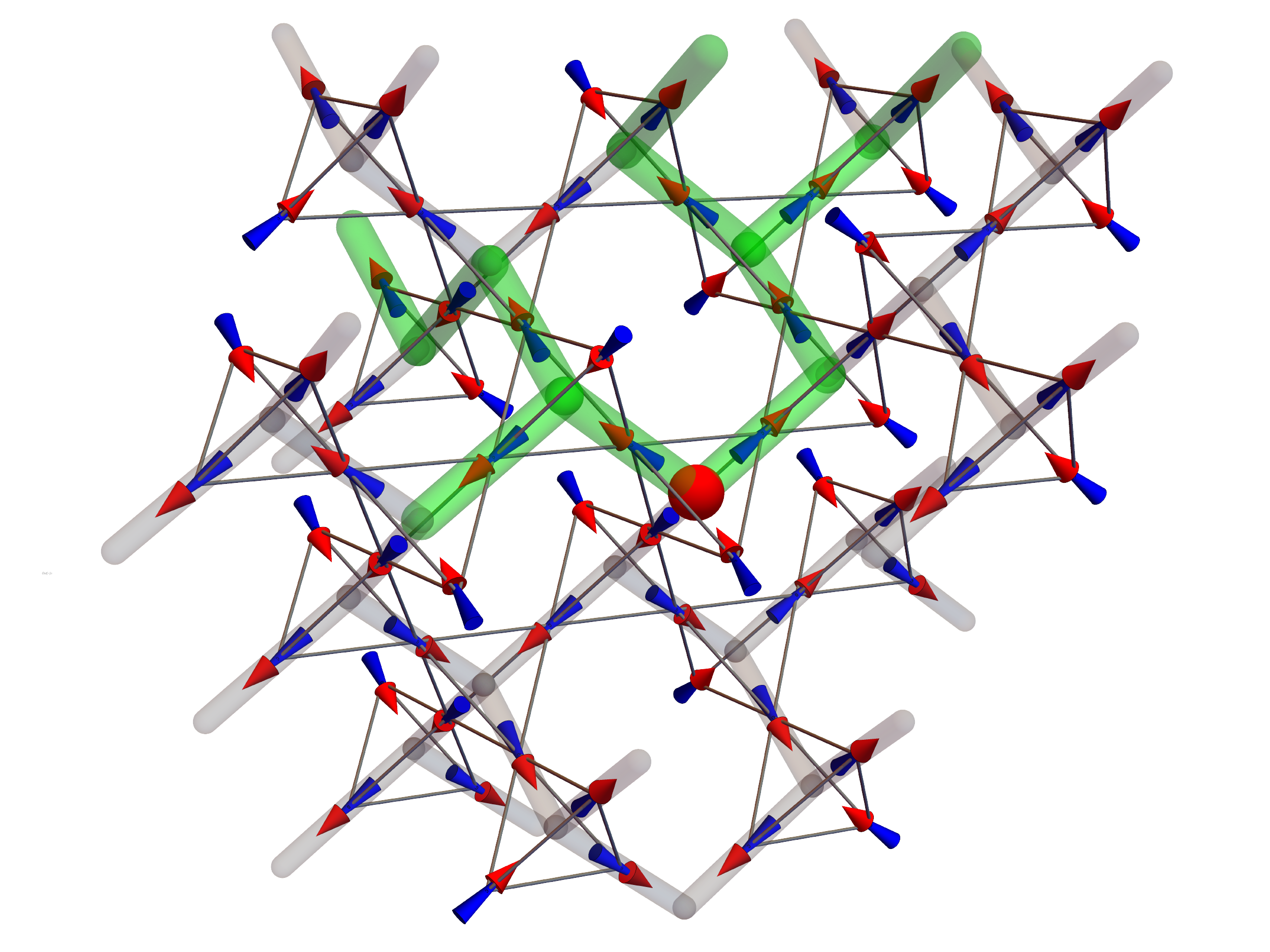}
    \caption{\label{fig:Pyrochlore} In spin ice, classical Ising-like moments reside on a pyrochlore lattice of corner-sharing tetrahedra. In each of the exponentially numerous ground states, two spins point into a tetrahedron, and two out. Violations of this ice rule take the form of magnetic monopoles (marked by a red sphere). These monopoles move on the diamond lattice formed by the centres of the tetrahedra. From a given position, monopoles can hop by flipping any one of the three majority spins (bonds highlighted in green and grey), but not the minority spin (unmarked bonds). Here we show parts of a spin ice configuration, including all tetrahedra that this monopole could reach within 3 hops. Some spins experience a vanishing transverse field \cite{tomasello2019}, and their dynamics is substantially suppressed. If moves through such spins are forbidden, the monopole is restricted to move on the paths highlighted in green – these form the emergent dynamical fractal.}
\end{figure}

The current intense research efforts on the behaviour of topological matter, besides unearthing many exciting phenomena, attempt to yield an understanding of these systems on the same level of both generality and detail as is available for conventional systems~\cite{chaikin_lubensky_1995,moessner_moore_2021}. One particular frontier concerns the {\it dynamical} properties, especially of topological systems that host exotic `fractionalised' excitations such as Laughlin quasiparticles with anyonic statistics in the quantum Hall effect~\cite{stern2008}, or emergent magnetic monopoles in the topological spin liquid known as spin ice~\cite{udagawa2021spin} (see Fig.~\ref{fig:Pyrochlore}). 
Indeed, the dynamical behaviour of the latter has been an enigma since its discovery~\cite{snyder2001,matsuhira2001,snyder2004}. Most recently, as well as strikingly, ultrasensitive low-temperature SQUID experiments have thrown up a new puzzle: the magnetic noise spectral density exhibits an {\it anomalous power law} as a function of frequency, with the low temperature exponent $\alpha\approx1.5$ deviating strongly from the well-known $\alpha=2$ of a paramagnet~\cite{dusad2019,samarakoon2022} (see also Ref.~\cite{kirschner2018,watson2019,Nisoli2021}). Within the -- generally successful -- framework of what we call the `standard model' of spin ice dynamics~\cite{ryzhkin2005,jaubert2009}, this behaviour cannot be accounted for using broadly accepted model Hamiltonian parameters~\cite{samarakoon2022}. 

%
%
%
%

In this work, we provide the missing ingredient: besides the constraints imposed by the emergent gauge field in spin ice, there is a further dynamical bottleneck on account of the local (transverse) field distribution, which suppresses the dynamics of another quarter of the spins~\cite{tomasello2019}. Both restrictions reflect the random yet correlated orientation of the spins in the spin ice ground states, and force the monopoles to move on an effectively disordered, slowly-time-evolving cluster in real space even in the absence of quenched disorder.

Crucially, this cluster is close to a percolation transition. It therefore exhibits a well-developed fractal structure on short and intermediate length scales, which we characterise in detail. We show that it is through hosting monopole motion that the fractal structure bequeathes anomalous exponents to the magnetic noise. Our numerical modelling of this process allows us to quantitatively reproduce the experimental noise curves, with only a single global fitting parameter for a microscopic timescale. In the process, we also shed light onto a further, long-standing puzzle in spin ice: the steeper than expected rise of the macroscopic relaxation time upon cooling. Our theory explains this phenomenon naturally, in a clean (i.e., stoichiometric and uniform) system, as a reflection of the sparseness and structure of the dynamical fractal. 

Although fractals have been found or artificially constructed in a range of systems such as porous materials~\cite{yu2008}, polymers~\cite{Newkome2006}, synthetic atom lattices~\cite{kempkes2019}, molecular systems~\cite{rothemund2004}, and the reciprocal space of quasi-crystals~\cite{viebahn2019}, we are not aware of any previous examples of fractals in clean stoichiometric bulk crystals. Notably, the fractal geometry in spin ice influences the dynamics in a qualitative and experimentally observable way while leaving no signatures in the thermodynamics
-- which is presumably why it has eluded discovery for so long.

\section{Model}
Spin ice is a topological magnet~\cite{Castelnovo_12,udagawa2021spin} with fractionalised quasiparticles in the form of mobile magnetic monopoles~\cite{Castelnovo2008}, as illustrated in Fig.~\ref{fig:Pyrochlore}. 
The low-temperature behaviour of spin ice in and out of equilibrium can largely be recast in terms of the dynamics of a dilute gas of monopoles~\cite{ryzhkin2005}, and how they interact with the background spin configuration. Specifically, the widely used `standard model' (SM) of incoherent spin ice dynamics forbids spin flips which create -- rather than hop -- monopoles (Fig.~\ref{fig:Pyrochlore}). For a monopole in a tetrahedron, this constraint systematically blocks one direction out of four~\cite{jaubert2009}. 

This model has successfully described important features of the dynamics -- especially the exponentially divergent relaxation time at low temperatures. However, it has failed to account for the large energy scale in the leading exponential growth of this time scale~\cite{matsuhira2011, pomaranski2013}, and it has in particular been challenged by susceptibility \cite{quilliam2011,yaraskavitch2012,revell2013} and anomalous magnetic noise experiments~\cite{dusad2019,samarakoon2022}. These puzzles and their resolution are discussed below.

Our new `beyond the standard' model (bSM) incorporates the observation that the internal field distribution on spins across which monopoles hop is peculiarly bimodal~\cite{tomasello2019}. In particular, one-third of the flippable spins experience a near-vanishing transverse field (see App.~\ref{app:MethodsA}). We model these spins as flipping at a lower rate $1/\tau_{\mathrm{slow}}$.
By contrast, the other spins experience a finite transverse field and flip at some reference rate $1/\tau_{\mathrm{fast}}$. 
(In the SM, all spins attempt to flip at a single rate $1/\tau_0$.) 
We take all $\tau$'s to be independent of temperature. 
The scale $\tau_{\mathrm{fast}}$ defines our unit of time, and -- as we will see -- it is in fact the only fitting parameter in our analysis. 

Regarding the interaction parameters, we use a model Hamiltonian, ${\cal H}_{\mathrm{OP}}$, an extension of the conventional dipolar spin ice Hamiltonian that was previously obtained from a combined fit to neutron scattering, magnetic susceptibility, and specific heat measurements~\cite{samarakoon2020}. It comprises long-range dipolar interactions and first, second, and third nearest-neighbour exchange terms. More details are given in App.~\ref{app:Hamiltonians} (where a comparison is drawn to the maximally simple nearest-neighbour spin ice model). 


\section{Results}
The left panel of Fig.~\ref{fig:ExpComp} shows the magnetic noise measured using SQUID magnetometry on a single crystal of DTO~\cite{samarakoon2022} in the temperature range $0.64~{\rm K} \leq T\leq 1.04~{\rm K}$, high enough to be above the 'freezing' of spin ice~\cite{snyder2004}, and low enough for monopoles to be sparse,  weakly-interacting quasiparticles. The noise is expressed in terms of the power spectral density (PSD), $S(\nu)$, defined as the temporal Fourier transform of the magnetisation, $M(t)$, autocorrelation function: $S(\nu) = {\cal F} \left[\langle M(0) M(t)\rangle \right]$.

Comparison with our simulations using ${\cal H}_{\mathrm{OP}}$ shows that bSM dynamics reproduces experiments over four orders of magnitude in frequency and six orders of magnitude in noise power (Fig.~\ref{fig:ExpComp}). The only fitting parameter is $\tau_{\mathrm{fast}}=85$~$\mu$s (we estimate $\tau_{\mathrm{slow}}/\tau_{\mathrm{fast}} \gtrsim 10^3$ (see App.~\ref{app:MethodsA}). This is indistinguishable from $\tau_{\mathrm{slow}}=\infty$ in these plots). By comparison, SM dynamics with fitting parameter $\tau_0=200$~$\mu$s is unable to describe the experimental data at low temperature.

%
%
\onecolumngrid
\begin{center}
\begin{figure}[ht!]
    \centering
    \includegraphics[width=0.49\textwidth]{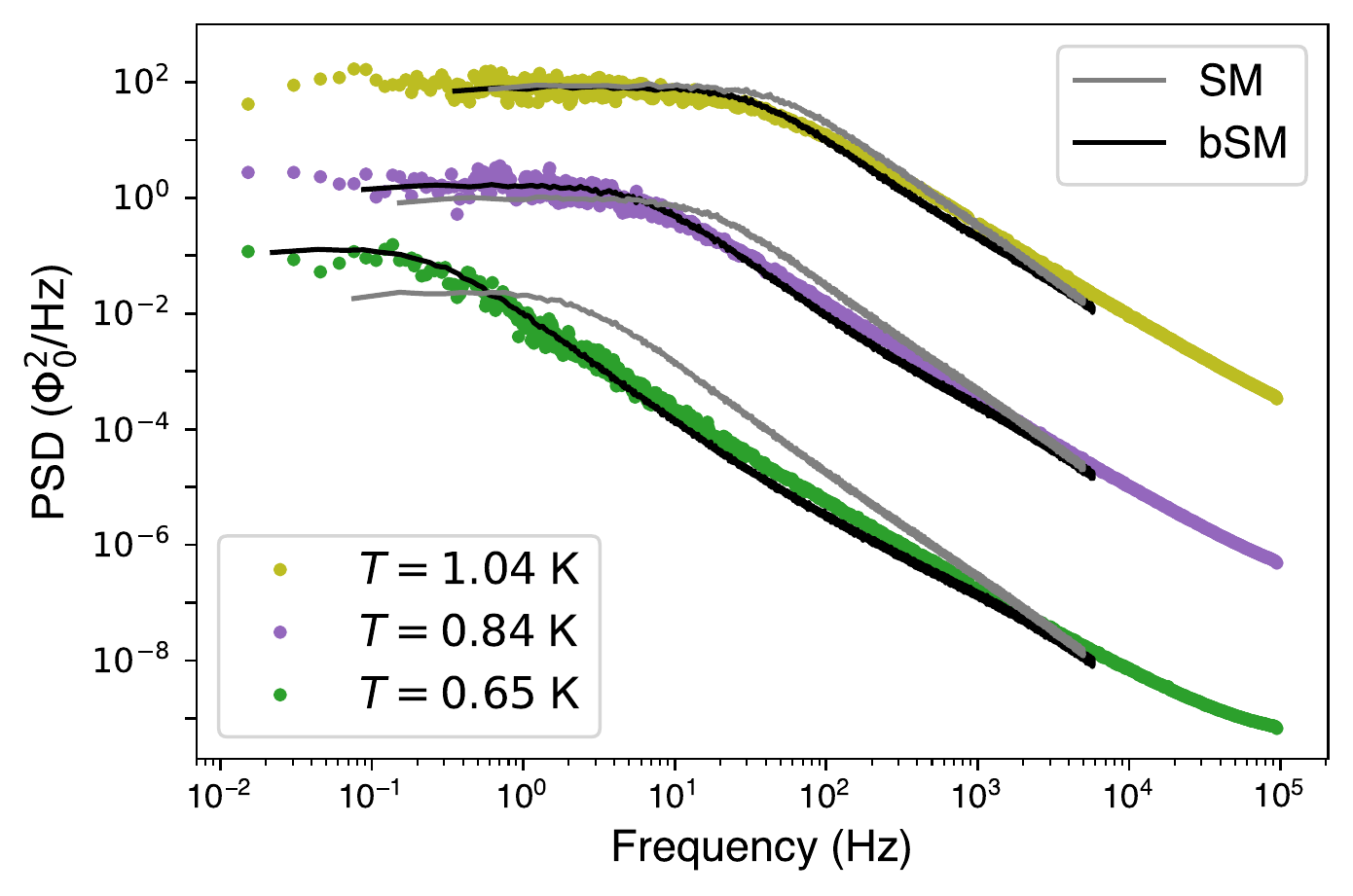}
    \includegraphics[width=0.49\textwidth]{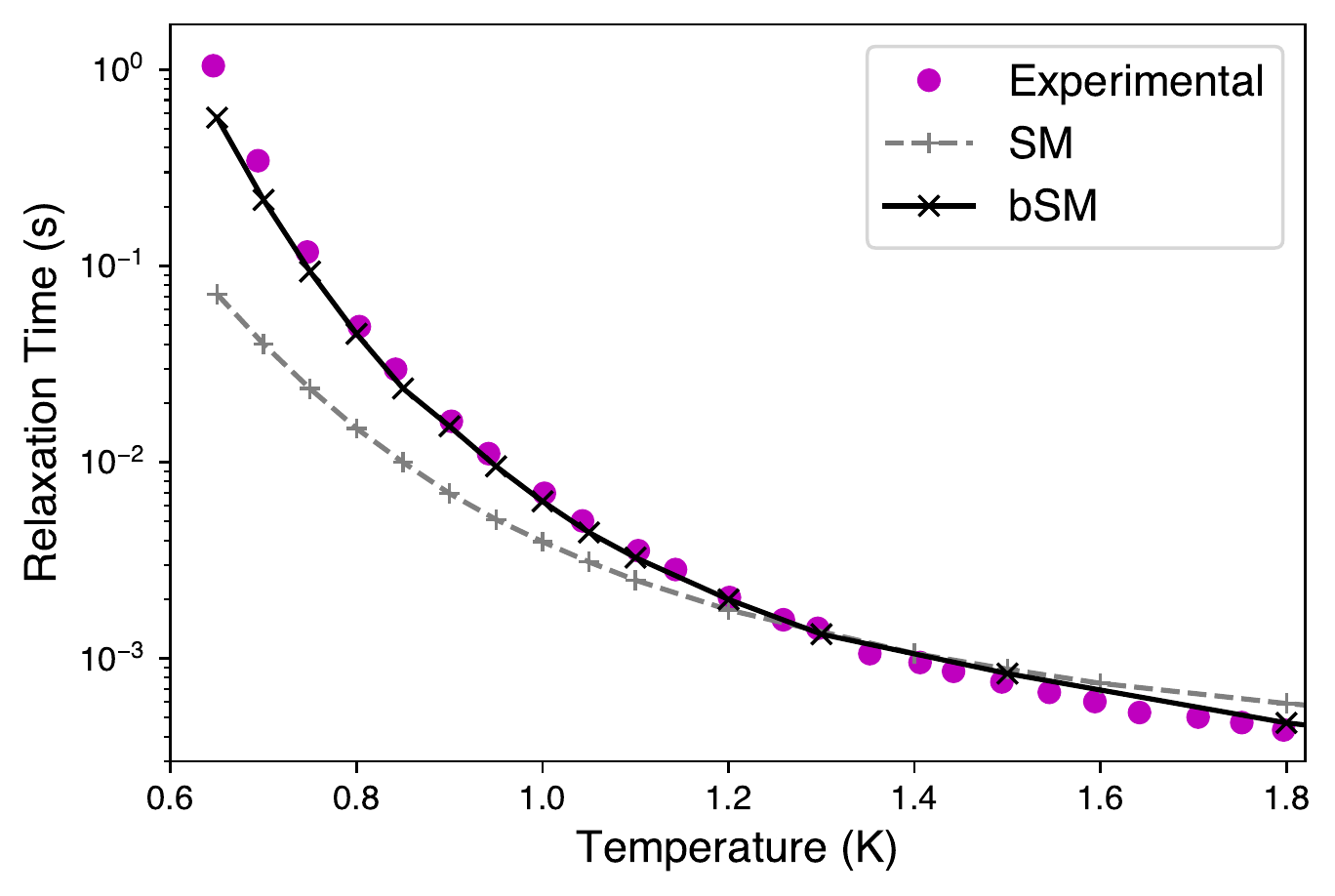}
    \caption{\label{fig:ExpComp}
    Left panel: Power spectral density (PSD) of magnetisation fluctuations extracted from SQUID measurements on Dy$_2$Ti$_2$O$_7$ (filled circles) \cite{samarakoon2022}. Monte Carlo results for ${\cal H}_{\mathrm{OP}}$ with SM dynamics (grey) and bSM dynamics (black) are shown with solid lines. The curves have been shifted vertically (see App.~\ref{app:MethodsA}). An overall time scale factor, $\tau_{\mathrm{fast}} = 85\ \mu$s and $\tau_0 = 200\ \mu$s, was applied for the bSM and SM results respectively to visually match the experimental data, with $\tau_{\mathrm{slow}} = \infty$ for simplicity. We note that the deviation at high frequencies is a known feature of Monte Carlo dynamics, and that other minor discrepancies are to be expected in a model with a necessarily sharper parameter distribution than the experimental system (see App.~\ref{app:MethodsA}).
    Right panel: Relaxation times $\tau$ were extracted from fits of the PSD curves to the function $A\left[1 + (2\pi f \tau)^\alpha \right]^{-1}$ for both experimental and numerical data.}
\end{figure}\ 
\end{center}\
\twocolumngrid\ 
%
%

The extracted relaxation time from the bSM (right panel of Fig.~\ref{fig:ExpComp}) also agrees remarkably well with experiment~\cite{samarakoon2022}, especially when contrasted with the SM, which yields much too short a relaxation time at low temperature. This has been a puzzle in the community for many years~\cite{matsuhira2001,snyder2004,ryzhkin2005,Jaubert2011,matsuhira2011,yaraskavitch2012,bramwell2012,revell2013,raban2022}: in a gas of freely moving monopoles, the relaxation time of the magnetisation scales with their inverse density~\cite{ryzhkin2005}, $1/\rho\sim\exp(\Delta_m/T)$, set by the energy cost $\Delta_m$ of an isolated monopole. 
Increasing the energy in the Arrhenius law to $\Delta_\tau > \Delta_m$ is largely precluded by basic statistical mechanics, whereas estimates suggest that a $\Delta_\tau$ in excess of twice $\Delta_m$ is actually required to fit the experimental growth of the relaxation time \cite{Jaubert2011}. Previous theories of the steep rise of the relaxation time upon cooling invoked {\it extrinsic} contributions due to open boundary effects, disorder, and an autonomously temperature-dependent microscopic time scale~\cite{revell2013,yaraskavitch2012,sala2014,paulsen2014,paulsen2016,edberg2022}; the identification of an {\it intrinsic} mechanism leading to a parametrically faster growth of the relaxation time than $1/\rho$ has been lacking.
\newpage

\begin{figure}[ht!]
    \centering
    \includegraphics[width=0.48\textwidth]{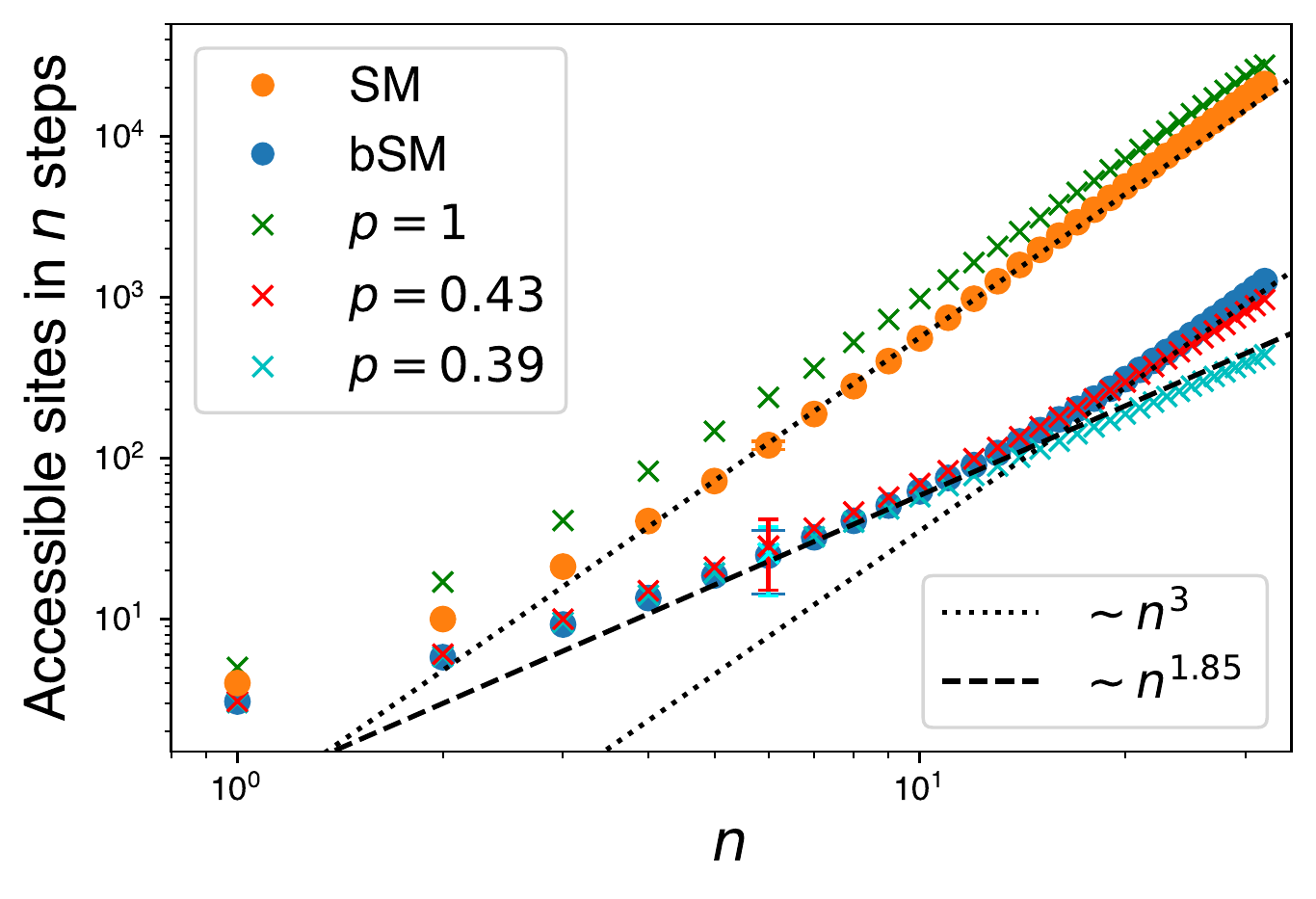}\\
    \includegraphics[width=0.48\textwidth]{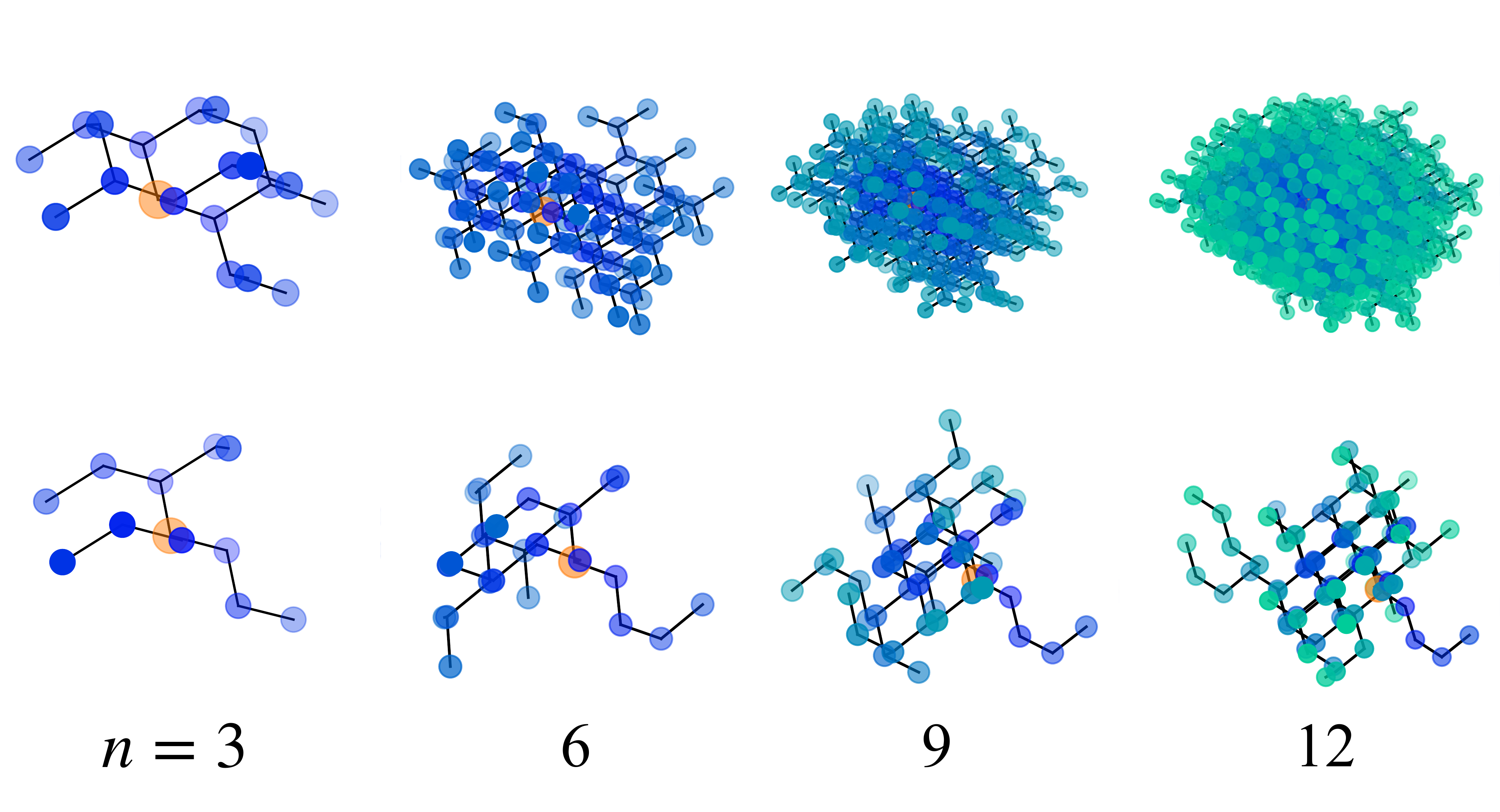}
    \caption{\label{fig:SitesReached}
    Top panel: Number of sites that a monopole can visit within a given number of steps $n$, the so-called chemical distance, with SM (orange circles) and bSM (blue circles) dynamics. The corresponding results for a walker on the diamond lattice at various bond filling fractions are shown by green, red, and cyan crosses. In the bSM case, there is a crossover between fractal $\sim n^{1.85}$ and standard $n^3$ scaling occurring at $n_{cr} \approx 14$, when the monopole has travelled a distance of the order of the correlation length. Within the fractal regime, i.e. up to $n_{cr}$, the monopole can access approximately 130 sites. The error bars, shown only for $n = 6$ for simplicity, reflect the finite width of the distribution and are not due to sampling (the statistical uncertainty is smaller than the size of the symbols).
    Bottom panel: Sites that a specific monopole, starting at the site marked in orange, can reach with SM dynamics (upper row) and bSM dynamics (lower row) within $n$ steps. The colour of the sites shifts from blue to green with the number of steps away from the starting site. The fractal cluster is much sparser, underpinning the rapidly diverging relaxation time (Fig.~\ref{fig:ExpComp}).}
\end{figure}

To explain the anomalous noise and susceptibility, and the corresponding strongly diverging relaxation time, we turn to what is perhaps the most remarkable discovery of this work. An isolated monopole in spin ice with bSM dynamics has between zero and three choices of sites to move to, with the statistical average being two. Linking the sites reachable by successive monopole hops, as in Fig.~\ref{fig:SitesReached}, yields a {\it fractal cluster}! Its fractal exponents are then picked up in the experimental anomalous noise signal, thus altering the relaxation properties of the system.

In more detail, to understand the fluctuations of the magnetisation, we need to analyse the statistical properties of the monopole motion, as (i) their motion proceeds via flipping of spins and (ii) they are the natural (sparse, weakly interacting) quasiparticles in the regime $T\lesssim 1$~K. 
The motion of the monopoles takes place on a `dynamical cluster' in real space, defined by excluding the spins which are not flippable energetically, or because of  a  small local transverse field. 

%
%
%
%
For $\tau_{\mathrm{slow}} =\infty$, this defines a percolation problem {\it close to the critical point}, so that the fractal structure is visible on small and intermediate scales (Fig.~\ref{fig:SitesReached}). In this regime, the number of sites the structure contains grows anomalously slowly with the chemical distance $n$ -- i.e., the minimum number of steps on the lattice needed to join two sites. We find that the growth agrees very well with $n^{1.85}$, the predicted exponent from critical percolation in $d=3$~\cite{Havlin1984,stauffer2018} (see Fig.~\ref{fig:SitesReached}), in contrast to the conventional $n^{d}$. 
Note the sparseness of the fractal: up to $n_{cr}=14$, it contains only about 130 of the 2071 sites available on the diamond lattice. Around $n_{cr} $, the system crosses over to conventional three-dimensional behaviour at longer lengthscales. Although our dynamical rules yield a non-standard correlated percolation problem, this does not impact its critical behaviour, and the properties of the cluster are in fact accurately reproduced by a random walker on a random percolation cluster on the diamond lattice at filling fraction $p=0.43$, only slightly above its critical value $p_c\approx0.39$~\cite{xu2014} (see App.~\ref{app:SuppText}).

Our discussion so far has largely been based on the behaviour of a single monopole. Beyond this, the motion of other monopoles can change the local spin and transverse field configurations, thereby endowing the percolation cluster with a slow dynamics of its own (see App.~\ref{app:SuppText}). A higher monopole density (alongside a finite $\tau_{\mathrm{slow}}$) therefore diminishes the crossover value $n_{cr}$.

It is the crossover in real space that terminates the anomalous regime in the PSD towards low frequencies in the time domain. At yet lower frequencies, a plateau appears on account of the trajectories of different monopoles overlapping. From the theory of random walks on percolation clusters above the percolation threshold~\cite{deGennes1976,Havlin1987,hughes2021}, the PSD of a monopole should show anomalous decay $S\sim \nu^{-(1+\sigma)}$ at high frequency and conventional decay $S\sim \nu^{-2}$ at low frequencies. Exact enumeration in the random percolation problem was previously used to obtain $\sigma=0.50 \pm 0.01$ in $d=3$~\cite{eduardo1990}. This largely explains the experimentally observed anomalous power law which hovers near $\nu^{-1.5}$ in the regime under consideration~\cite{samarakoon2022}.


\section{\label{Sec:conclusion}
Discussion
         }
As is so often the case with discoveries in condensed matter physics, their beauty derives not only from the notability of the new phenomena in itself, but also from the often unexpected ingredients of the explanation, and the way they snugly fit together.

From the point of view of percolation, we have discovered a purely dynamically-generated fractal object in a uniform, stoichiometric, disorder-free bulk crystal. 
Its existence is predicated on the emergence of point-like mobile objects -- the magnetic monopoles -- in a three-dimensional topological spin liquid which remains fluctuating down to low temperatures \cite{Castelnovo2008}.
The monopole motion is subject in turn to a twofold set of constraints. One, imposed by the emergent gauge field represented by the spin background, reflects the large-scale topological nature of the spin ice state. The other is purely microscopic in origin, resulting from an interplay of the geometric spin arrangement and the short-range statistical spin correlations imposed by the ice rules. It is the combination of these constraints that makes the monopoles move on a fractal structure. This in turn is the origin of the previously puzzling anomalous magnetic noise and rapidly diverging relaxation time. Conversely, this highly non-trivial phenomenology provides a rare validation of a concrete model of the microscopic dynamics, in particular the bimodal distribution of local transverse fields.

Most remarkably, this is all accessible by probing magnetisation response and fluctuations, specifically using AC-susceptibility or noise measurements, whose {\it frequency} dependence reflects {\it spatial} information, linked by the (sub-)diffusive motion of the monopoles. The emergence of a dynamical fractal in a disorder-free crystal presents a novel mechanism for the existence of anomalous noise, a subject which has been studied in many other contexts, within materials science~\cite{aliev2007,balandin2013,najafi2021,costanzi2017} and elsewhere~\cite{voss1975,press1978,varotsos2013,vernotte2015}.
Our work provides the first bulk magnetic measurement of a fractal.

Some of the above ingredients can be manipulated in a controlled manner, all of which provide promising avenues for future experiments. Trivially, different spin ice compounds have different interaction parameters, resulting in broadly different transverse field distributions. Moreover, the microscopic dynamics can also be altered by applying uniaxial strain~\cite{pili2021} or via the crystal field scheme, by switching from (Kramers) Dy- to, say, (non-Kramers) Ho-based spin ice. For example, HTO is predicted to have a much smaller ratio $\tau_{\mathrm{slow}}/\tau_{\mathrm{fast}}$~\cite{tomasello2019}. 

Note that our effective percolation problem has turned out to be at a sweet spot near criticality -- reducing constraints would eliminate the anomalous nature of the signal, whereas increasing them would likely eliminate equilibration. Indeed, it will be interesting to see if doing so (e.g., by the controlled introduction of some form of quenched or dynamical disorder) may shed light on the `glassy' physics below $650$~mK~\cite{bramwell2001}. 

Needless to say, the advent of highly-tunable NISQ platforms opens up an entirely new set of directions, especially in two dimensions~\cite{King2021,Stern2021}, where anomalous noise has already been seen in the classical nanomagnetic artificial spin ice~\cite{Goryca2021}. This includes questions related to quantum diffusion of monopoles and the role of increasingly coherent many-body quantum dynamics. The latter could alter the noise in a characteristic way (see App.~\ref{app:SuppText}) and therefore be used as evidence for the presence of so-called ring exchange processes in candidate quantum spin ice compounds. 

In closing, we note that it was thanks to the wealth of information accumulated over 25 years of work by the magnetism community~\cite{bramwell2001,udagawa2021spin} that we were able to analyse the cooperative dynamics of spin ice on the level of detail required for discovering the dynamical fractal. This underlines the value of well-characterised model systems for driving the discovery of striking yet subtle phenomena. 


\section*{Acknowledgments}
We express our gratitude to Cristiano Nisoli, Peter Schiffer, Shivaji L. Sondhi and Shu Zhang for useful discussions.
This work was supported in part by the Engineering and Physical Sciences (EPSRC) grants No.~EP/P034616/1 and EP/T028580/1 (CC and JNH), the Deutsche Forschungsgemeinschaft under grants SFB 1143 (project-id 247310070) (RM and JNH), and the cluster of excellence ct.qmat (EXC 2147, project-id 390858490) (RM and JNH).
The work of DAT was supported by the U.S. Department of Energy, Office of Science, Office of Basic Energy Sciences, under Award Number DE-SC0022311.
Part of this work was carried out within the framework of a Max Planck independent research group on strongly correlated systems.


\bibliography{references}

\clearpage 
\newpage

\appendix

\section{Methods
\label{app:MethodsA}}

\subsection{Model Hamiltonians
\label{app:Hamiltonians}}
We use two different Hamiltonians for the spin-spin interactions. The first one was obtained as a combined fit to neutron scattering, magnetic susceptibility, and specific heat measurements on Dy$_2$Ti$_2$O$_7$ \cite{samarakoon2020}. It consists of long-range dipolar interactions and short-range exchange between first, second, and third-neighbour spins:
\begin{eqnarray}
\label{eq:H_comp}
{\cal H}_{\rm{OP}}\   \!\!\! &=& \!\!\! \  Da^3 \sum_{i<j}\left[
\frac{\Vec{S}_i \cdot \Vec{S}_j}{r_{ij}^3} - \frac{3 \left( \Vec{S}_i \cdot \Vec{r}_{ij}\right) \left( \Vec{S}_j \cdot \Vec{r}_{ij}\right)}{r_{ij}^5}  \right]
\nonumber \\ 
\!\!\! &+& \!\!\!  \  J_1\sum_{\left<i,j\right>}  \Vec{S}_i \cdot \Vec{S}_j + J_2 \sum_{\left<i,j\right>_{2}}  \Vec{S}_i \cdot \Vec{S}_j 
\nonumber \\
\!\!\! &+& \!\!\! J_3\sum_{\left<i,j\right>_3}  \Vec{S}_i \cdot \Vec{S}_j  
+ J_3'\sum_{\left<i,j\right>_{3'}}  \Vec{S}_i \cdot \Vec{S}_j
\, . 
\end{eqnarray}
The dipolar interaction strength is $D=1.3224 \ \mathrm{K}/a^3$, and the exchange strengths are $J_1=3.41$~K, $J_2=0.0$~K, $J_3=-0.00466$~K, and $J_3'=0.0439$~K~\cite{samarakoon2022}. 

For reference, we contrast this with the simplest possible model: the nearest-neighbour Hamiltonian, 
\begin{equation}
    \label{eq:H_NN}
    {\cal H}_{\rm{NN}} = -J_{\rm{eff}}\sum_{\left<i,j\right>}\Vec{S}_i \cdot \Vec{S}_j \, 
\end{equation}
that has an effective exchange parameter $J_{\mathrm{eff}}$. In our work $J_{\mathrm{eff}}$ was set to $5.7$~K, chosen so that the monopole number densities of ${\cal H}_{\mathrm{OP}}$ and ${\cal H}_{\mathrm{NN}}$ approximately match at $T=0.75$~K. 

In regards to the phenomena of interest in our work -- namely the magnetic noise, the relaxation time scale, and the emergent dynamical fractal-- we find that ${\cal H}_{\rm{OP}}$ and ${\cal H}_{\rm{NN}}$ give qualitatively similar results. For this reason, part of the analysis presented in the appendices uses ${\cal H}_{\rm{NN}}$ for computational convenience and to access larger system sizes. 


\subsection{Transverse fields and spin flipping rates
\label{app:transverse fields}}
\begin{figure}
    \centering
    \includegraphics[width=1.0\columnwidth]{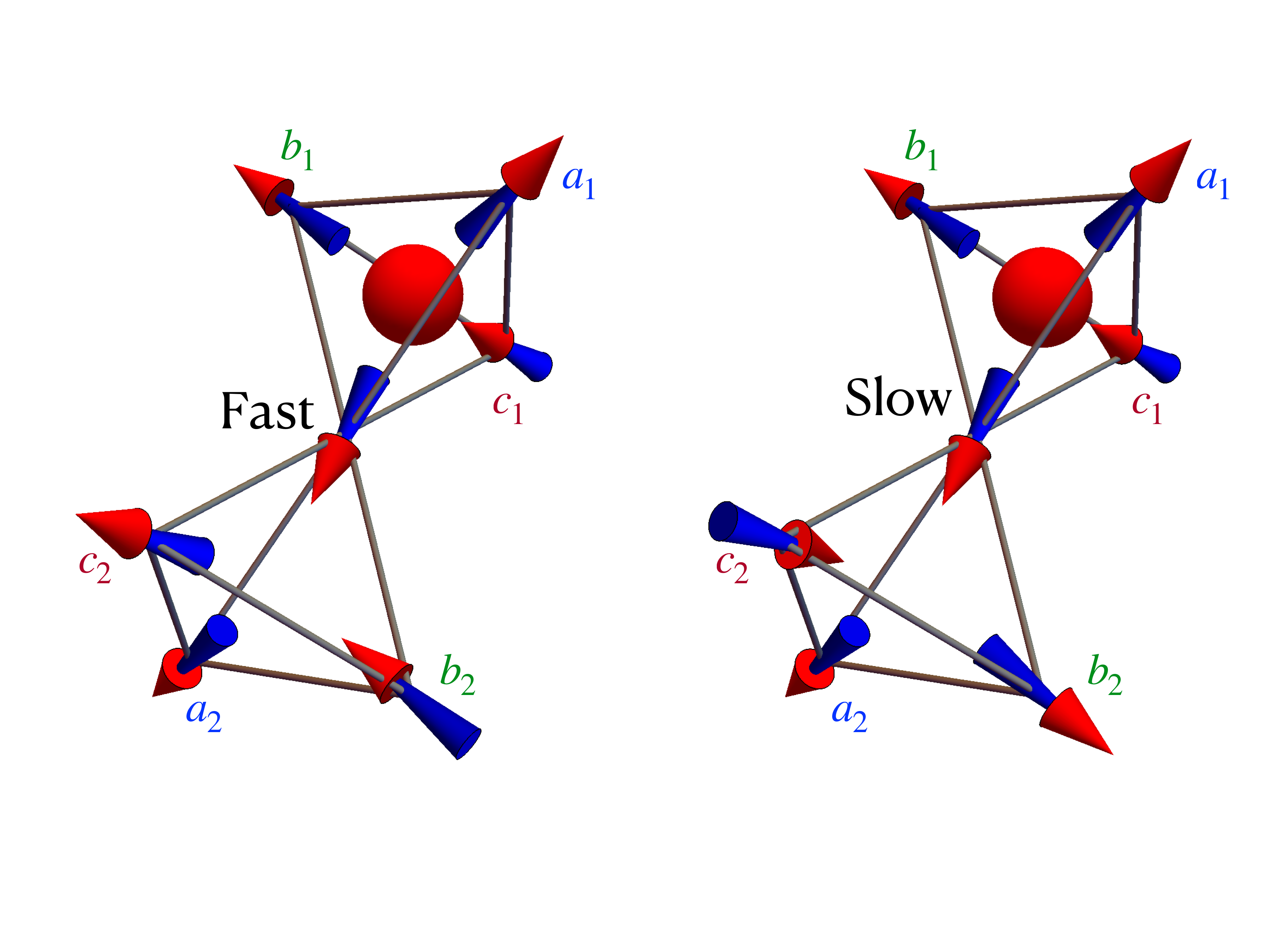}
    \caption{Two tetrahedra in spin ice with a single monopole. The easy axes of spins $a_1$ and $a_2$, $b_1$ and $b_2$, and $c_1$ and $c_2$ are parallel to one another. The longitudinal field on the central spin from its six nearest-neighbours vanishes in both cases and the central spin can therefore flip without energy penalty, enabling the monopole to hop. In the left scenario, there is a non-zero effective transverse field from the nearest-neighbour spins, corresponding to $\tau_{\rm fast}$. In the configuration on the right, the effective transverse field vanishes and there are no matrix elements that enable the spin to flip~\cite{tomasello2019}, corresponding to $\tau_{\rm slow}$.} 
    \label{fig:fast&slow}
\end{figure}

In Ref.~\cite{tomasello2019} it was observed that spin ice Hamiltonians generally lead to a bimodal distribution of effective transverse fields at the site of a spin whose flip leads to the hop of a monopole (see Fig.~\ref{fig:fast&slow}). As explained in detail there, this arises because the six spins neighbouring a given site provide environments with two different symmetries. In the high-symmetry case (right panel of Fig.~\ref{fig:fast&slow}), which occurs in a third of the instances, the site is an inversion centre, leading to a vanishing net field due to the six neighbours. 
In the low-symmetry case, by contrast, it is only the longitudinal field component which vanishes, while the transverse one remains finite. Farther range interactions, of dipolar or exchange origin, are found to only introduce minor corrections. (Specifically to ${\cal H}_{\mathrm{OP}}$, the second neighbour exchange terms vanish, and the third neighbour ones are not expected to have a transverse component.) 

%
%
%
%

In the standard model (SM) of spin ice dynamics, this difference is not included and all spins are assumed to flip stochastically with an average characteristic rate $1/\tau_0$. In this work we consider instead `beyond the standard' model (bSM) dynamics, which attempts to flip spins at a slow rate $1/\tau_{\mathrm{slow}}$ if the local field has a vanishing transverse component, and at a fast rate $1/\tau_{\mathrm{fast}}$ otherwise. 
In all cases, the time scales are assumed to be independent of temperature. 


\subsection{Methods -- spin ice}
All numerical results were obtained using Monte Carlo simulations with the Metropolis algorithm, or variations thereof as explained below, with periodic boundary conditions. 

We denote the total number of spins in the system by $N$ (note that the number of spins on the pyrochlore lattice equals the number of bonds on the diamond lattice). A $16$ site cubic unit cell was used, and we considered systems of linear size $10$ ($N=16000$) for the thermal noise and diffusion measurements, and of linear size $16$ ($N=65536$) for the characterisation of the percolation cluster and measurements that involve a single mobile monopole (see below).

Spin ice configurations at different temperatures were prepared by simulated annealing from a high-temperature random configuration. The Ewald summation technique was used to include long-ranged dipolar interactions. Worm updates were used to ensure equilibration at the lowest temperatures, and when preparing the system for single mobile monopole simulations (more on this later). Worm updates consist of finding a chain of spins aligned head to tail, forming either a closed loop or terminating at a monopole at one end, and attempting to flip the corresponding spins simultaneously. The update is accepted with the customary Metropolis probability $P= \min{\left[1, \exp \left( \Delta E / T\right) \right]}$, where $\Delta E$ is the overall change in energy. If the update is accepted, it neither creates nor annihilates any monopoles. 

The dynamics of thermal systems was simulated using single spin flip updates. A random spin is selected and the energy cost $\Delta E$ associated with flipping this spin is calculated. In the case of bSM dynamics, the move is accepted with probability $P=G \, \min{\left[1, \exp \left( \Delta E / T\right) \right]}$, where $G=1$ if the spin flips with time scale $\tau_{\mathrm{fast}}$, and $G=\tau_{\mathrm{fast}} / \tau_{\mathrm{slow}}$ if the spin flips with time scale $\tau_{\mathrm{slow}}$. 
Slow and fast time scales were determined based on the configuration of the six nearest-neighbouring spins, according to the discussion above -- see also Ref.~\cite{tomasello2019}. 

A Monte Carlo sweep consists of $N$ attempted updates, and one sweep corresponds to the time scale $\tau_0$ in SM dynamics, and to the time scale $\tau_{\mathrm{fast}}$ in bSM dynamics. In the monopole picture this allows for monopole creation, annihilation, and motion. Double monopoles are allowed, although they quickly become energetically excluded at low temperatures.

In the thermal simulations, we are also able to follow specific monopoles from creation to annihilation. When doing this, we first equilibrated the system and then tracked the monopoles created afterwards, up to the point when they are annihilated. 
From the measured trajectories, it is possible to extract properties such as the monopole lifetime or radius of gyration distributions. One can also compute the PSD of each individual monopole trajectory; however, as this requires more intensive simulations than to compute the PSD of the magnetisation of the entire system, we limited ourselves to consider only the higher portion of the frequency spectrum, $\nu > 10^{-3}/\tau_{\rm fast}$. 

\subsubsection{Single monopole simulations}
The dynamical magnetic properties of spin ice, in the temperature regime under consideration, are best understood in terms of the statistical properties of monopole motion. Monopole moves occur through spin flips, and magnetisation fluctuations are thus directly linked to monopole displacement. 

Single mobile monopole simulations are a good approximation of ${\cal H}_{\mathrm{NN}}$ at low temperatures (low monopole density), when one neglects monopole creation and annihilation events. The simulations were initiated from a random spin ice state with zero monopoles. A pair of monopoles was then generated, one of which was kept at a fixed location throughout the simulation (annihilation events between the pair were thereafter forbidden).
The dynamics is generated from updates consisting of randomly selecting one of the four spins neighbouring the mobile monopole, and flipping this if the spin (a) is a majority spin and (b) has a non-zero transverse field (for bSM dynamics only). Four such attempted updates correspond to one Monte Carlo sweep in the thermal simulations, and we therefore equate this to $\tau_0$ (SM) or $\tau_{\mathrm{fast}}$ (bSM). This assumes that the monopole can move at no energy cost, as is the case in the nearest-neighbour model. Before any measurements were taken, the mobile monopole was allowed to perform $6.6\times 10^7$ attempted moves to ensure that the positions of the mobile and static monopoles had become sufficiently uncorrelated.

\begin{figure}
    \centering
    \includegraphics[width=1.0\columnwidth]{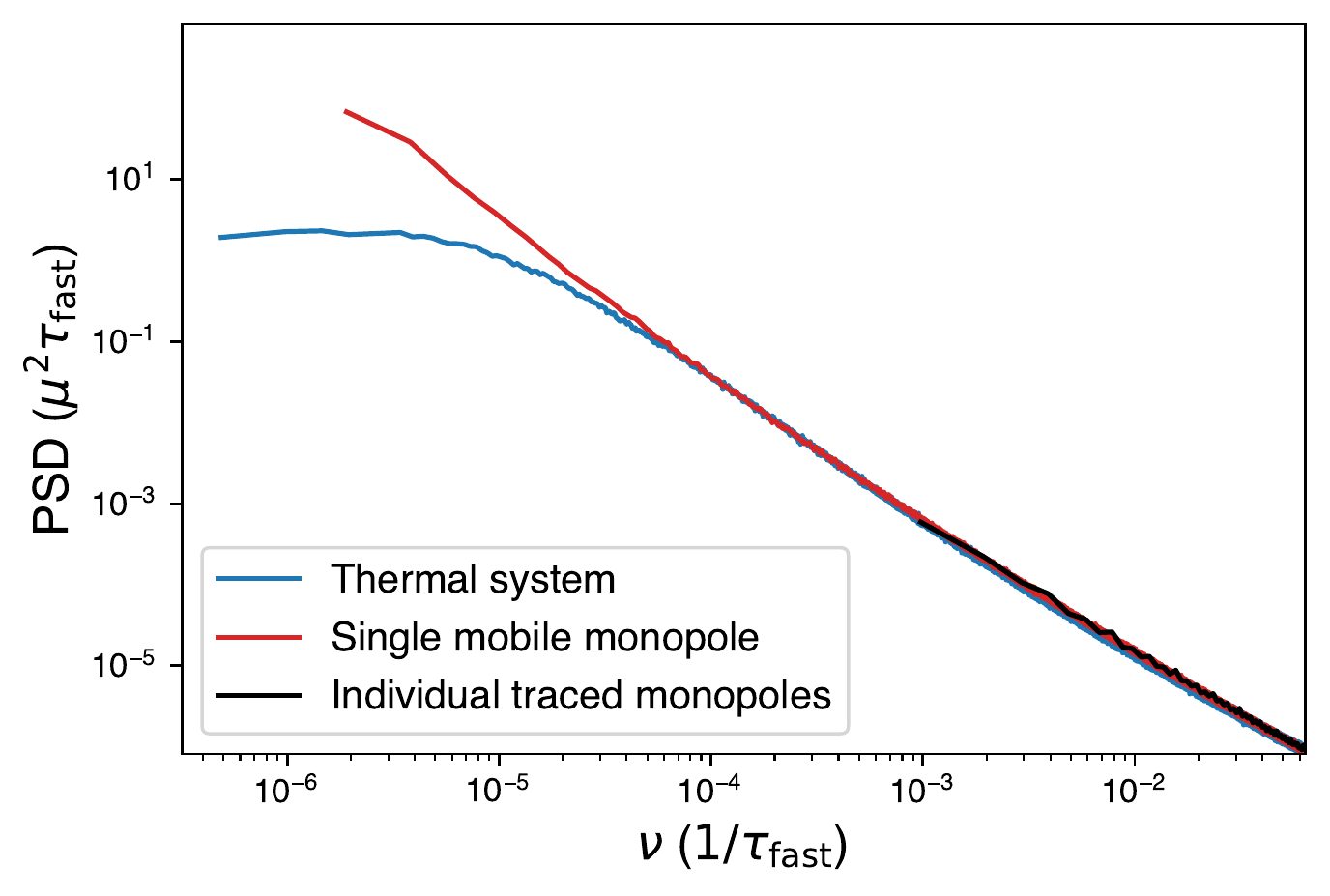}
    \caption{\label{fig:3modelComparison}
    Comparison of the PSD computed using Monte Carlo simulations of ${\cal H}_{\mathrm{NN}}$ at $T=0.6$~K (blue), using single mobile monopole simulations (red), and using the trajectories of individual monopoles in the Monte Carlo simulations (black, obtained only for frequencies $\nu \gtrsim 10^{-3}/\tau_{\rm fast}$). The red and black lines have been rescaled so that the three cases match in magnitude.}
\end{figure}

To control that the single monopole simulations do indeed reproduce the magnetic noise of the thermal system at low temperatures, we compared the PSDs obtained from the thermal simulations, single mobile monopole simulations, and individual monopole trajectories in the thermal simulations (Fig.~\ref{fig:3modelComparison}). After rescaling them to match in magnitude, the three cases agree well, indicating that the low frequency noise can indeed be understood from the motion of individual monopoles.

%
%
%
%


\subsection{Methods -- bond percolation}
Uncorrelated bond percolation configurations at different filling fractions were generated by randomly placing bonds on the diamond lattice. As a way to further mimic spin ice, a specific type of correlation was also considered, where the bonds are placed subject to the constraint that no vertex is connected to more than three bonds. This mimics the constraint that every site containing a single monopole in spin ice has one minority spin that the monopole cannot move through. The constraint was implemented by first creating an ordered bond configuration with fixed connectivity, ensuring that every vertex has exactly three bonds. The configuration was then scrambled by randomly applying dimer loop updates, and further bonds were subsequently removed at random to reach the required filling fraction. We find that this constraint increases the percolation threshold to $p\approx 0.41$, above which a structured percolation cluster is created. (Here the word \emph{structured} is used to refer to the extra correlations introduced by the constraint.) 

The random walk on the uncorrelated percolation cluster (PC) or the structured percolation cluster (SPC) works similarly to the single mobile monopole simulations. A walker is placed at a random vertex on the percolation cluster. The updates consist of randomly choosing one of the four bonds at that vertex and moving the walker along it, if the bond exists. This is the diffusion model commonly referred to as \quotes{the blind ant in the labyrinth}~\cite{deGennes1976,Havlin1987}. To enable comparison with bSM spin ice, we consider four such attempted updates to be equivalent to $\tau_{\mathrm{fast}}$.


\subsection{Connected clusters}
An adapted depth-first search~\cite{even2011} was used to identify clusters of connected sites. Following the standard algorithm, a monopole or walker is moved from an initial reference site to all sites it can reach within a specific chemical distance (the minimum number of steps on the lattice) from the reference site. In the case of monopoles, the spin configuration is updated as the search is performed, and backtracking from dead-ends or sites at the largest chemical distance must be performed explicitly to maintain the correct spin configuration.


\subsection{Power spectral density}
PSDs were calculated using Welch's method. The high frequency aliasing tail (namely, beyond the Nyquist frequency) has been cut off in the examples shown. Relaxation times $\tau$ were extracted by fitting the low frequency knee with the function
\begin{equation}
    f = A \frac{\tau}{1 + \left(\tau \nu \right)^\alpha}
    \, .
    \label{eq:Cole-Cole} 
\end{equation}


\subsection{Experimental and numerical units 
\label{app:compareExpandSim}}
The experimental magnetic noise data shown in Fig.~2 of the main text were recorded in units of $\Phi_0^2/{\rm Hz}$~\cite{samarakoon2022} ($\Phi_0 = h/2e \simeq 2.068 \cdot 10^{-15}$~Vs being the magnetic flux quantum). 

In our simulations we measure the magnetisation in units of the single spin magnetic moment strength $\mu$, which for Dy$_2$Ti$_2$O$_7$ is approximately $10\mu_B$ (Bohr magnetons) \cite{udagawa2021spin}. The natural unit of time is $\tau_{\mathrm{fast}}$ in simulations with bSM dynamics and the $\tau_0$ in simulations with the SM dynamics. When comparing the models below we use $\tau_0=\tau_{\mathrm{fast}}$. The PSD computed from simulations is measured in units of $\mu^2 \tau_{\mathrm{fast}}$. 

The numerical results in Fig.~2A in the main text have been scaled to match the experimental ones; unfortunately the latter depend on details such as the size of the squid and the portion of sample contributing to the magnetic flux signal, which carry too much uncertainty to be able to derive the appropriate rescaling factor from first principles. Based on visual inspection, factors of 850 and 795 respectively were used to rescale the SM and bSM numerical PSD curves. The same factor was used for all displayed temperatures. For visual clarity the PSD curves were shifted vertically by multiplicative factor $2^0$, $2^4$ and $2^8$ for temperature 0.65, 0.84, and 1.04~K respectively. 



\section{Supplementary Text
\label{app:SuppText}}


\subsection{Comparison between simulations and experiments}
The agreement between bSM simulations and experimental results (displayed in Fig.~2 of the main text) is quite remarkable, considering our model only has one global fitting parameter. There are however some discrepancies which we comment on here. 

Deviations at high frequencies are a known feature of simple Monte Carlo dynamics (see for example \cite{samarakoon2022}). Monte Carlo dynamics relies on the approximations that a single spin attempts to flip at uncorrelated times and that it on average attempts one flip in time $\tau_{0}$ (or $\tau_{\mathrm{fast}}$ in the bSM case); it models a Poissonian process. As a result of this, the PSD for frequencies $\nu > 1/\tau_{0}$ always decays as $\nu^{-2}$.
At the same time, the experimental signal at the highest frequencies is very small (for $T = 0.65$~K, it is about eight orders of magnitude below the low-frequency signal), and the relative upturn in the experimental curve there is likely also a signal-to-noise issue.

The theoretical model naturally has a sharper parameter distribution than the experimental system, and this leads to a signal with sharper crossovers between different regimes, which would account for why the numerical curve overshoots the experimental one somewhat at low frequencies, and then drops slightly below it at intermediate frequencies. Future work may help fine-tune the description and reduce these discrepancies; however, this will require a separate microscopic determination of any additional parameters (e.g., of the value of spin flip time scales like $\tau_{\mathrm{fast}}$), lest it be a mere fit improvement by increasing the number of adjustable parameters.


\subsection{\label{app:PC-SIcomparison}
Constrained spin ice as a correlated percolation problem}
\begin{figure}
    \centering
    \includegraphics[width=1.\columnwidth]{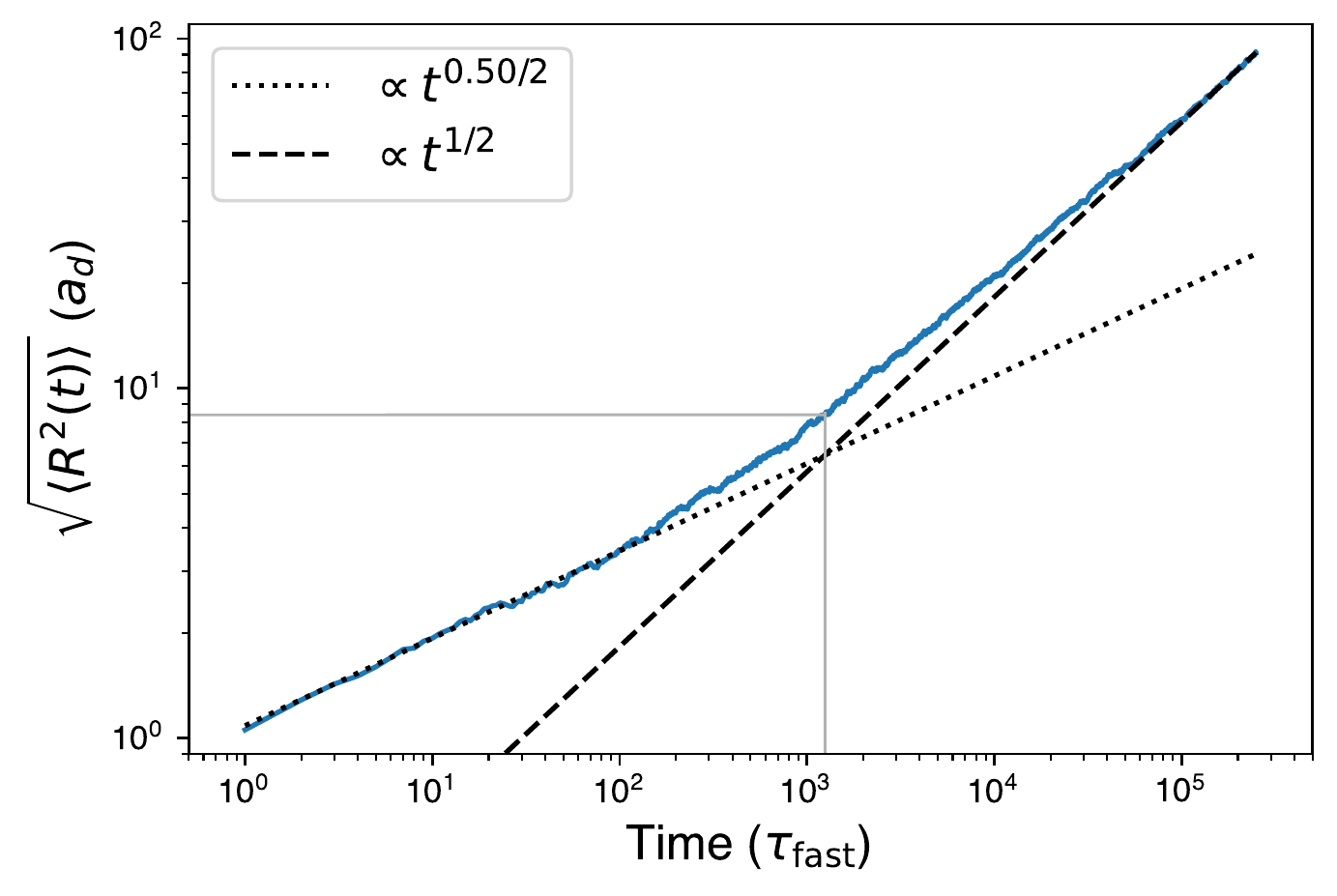}
    \caption{\label{fig:MSD}
    The root-mean-squared displacement of a monopole in bSM spin ice, measured in units of the diamond lattice nearest-neighbour distance $a_d$. From the crossover behaviour we can extract a typical time and length scale for the fractal behaviour, and we find that they are approximately $1200\tau_{\mathrm{fast}}$ and $8 a_d$ respectively.}
\end{figure}

The fractal nature of the clusters the monopoles move on arises from the proximity to the percolation threshold when one maps blocked spins onto missing bonds in a bond percolation problem. In the standard language of percolation theory \cite{stauffer2018}, the percolation cluster first appears at the percolation threshold $p_c$. For bond percolation on the diamond lattice, this has been numerically estimated at $p_c \approx 0.39$ \cite{xu2014}. At this point, the percolation cluster is self-similar on all length scales. If one increases the filling fraction beyond the threshold value, the percolation cluster remains fractal only on length scales shorter than the correlation length $\xi$, which decreases as the filling fraction increases. A random walker moving on the percolation cluster will move sub-diffusively on short length scales, with a mean-squared displacement $\langle R^2(t) \rangle \sim t^{\sigma}$ and $\sigma < 1$. For three-dimensional lattices, $\sigma=0.50 \pm 0.01$ has been computed using exact enumeration \cite{eduardo1990}. (Note that our $\sigma$ is directly related to $d_w$ in the notation of Ref.~\cite{eduardo1990}, with $\sigma = 2/d_w$ and $d_w$ defined by $\sqrt{\langle R^2(t) \rangle} \sim t^{1/d_w}$.). On scales greater than $\xi$ the motion of the walker becomes diffusive, $\langle R^2(t) \rangle \sim t$. This behaviour is observed for a monopole in bSM spin ice (Fig.~\ref{fig:MSD}). After a time of approximately $10^3\ \tau_{\mathrm{fast}}$, the crossover from sub-diffusive to diffusive behaviour occurs. The corresponding length scale of the crossover is approximately $8\ a_d$, where $a_d$ is the nearest-neighbour distance on the diamond lattice.

%
%
%
%

\begin{figure}
    \centering
    \includegraphics[width=1.\columnwidth]{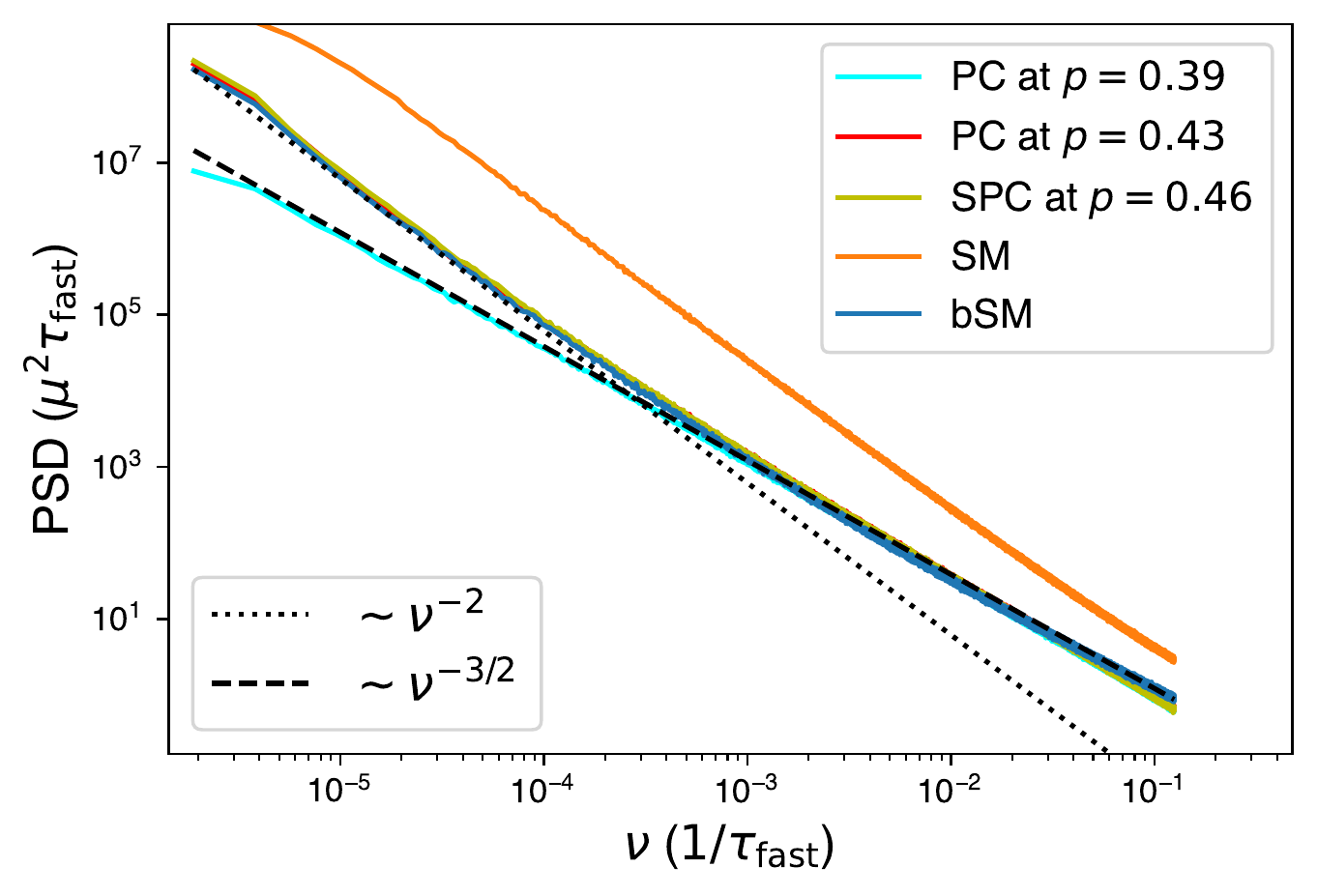}
    \includegraphics[width=1.\columnwidth]{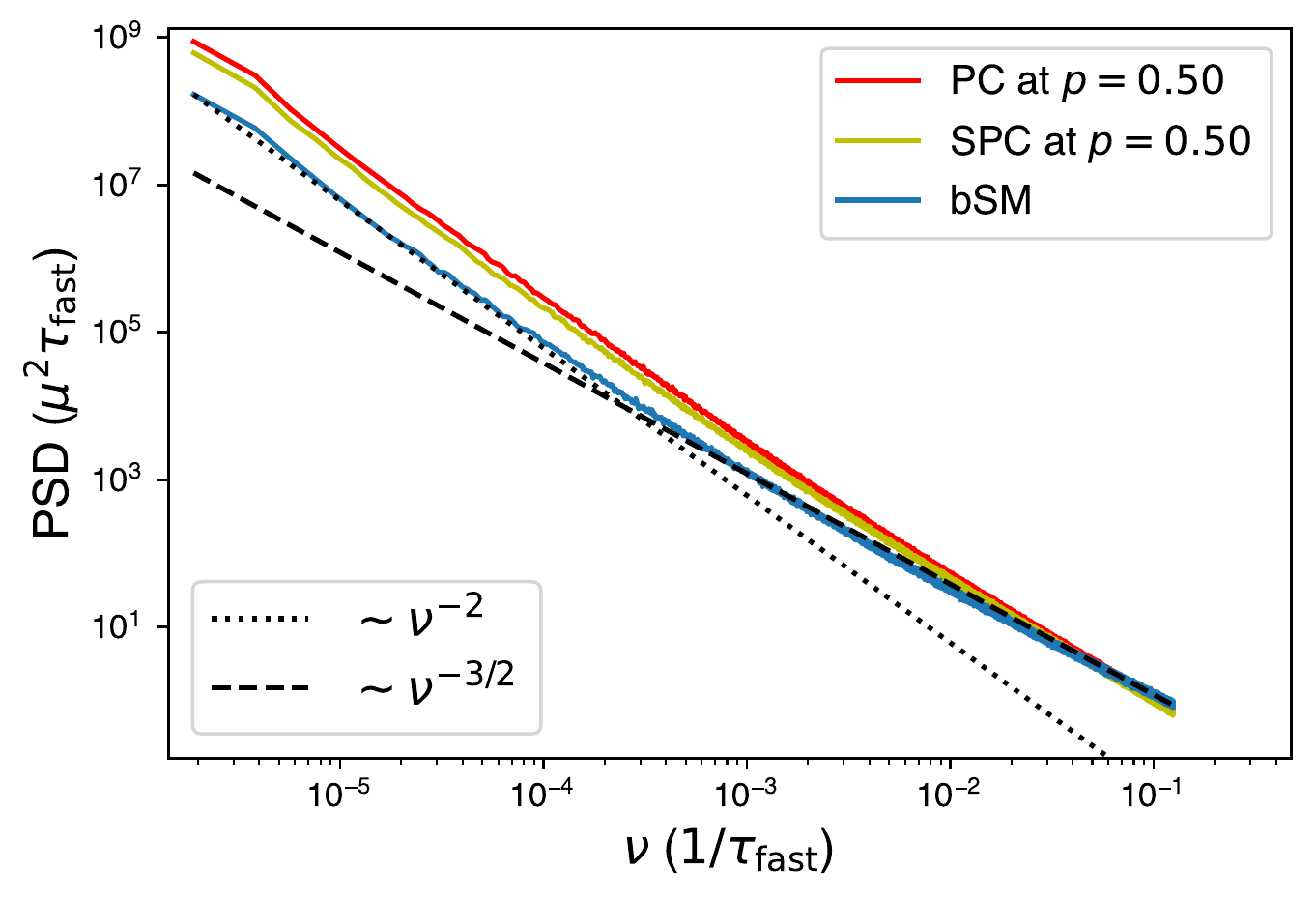}
    \caption{\label{fig:SI+PC}
    PSDs computed from Monte Carlo simulations of a random walk on the uncorrelated percolation cluster (PC) and on the structured percolation cluster (SPC) on the diamond lattice. The cyan line shows that the PSD indeed goes as $\nu^{-1.5}$ when $p=p_c$. The PSD computed for the position of a single mobile monopole with bSM nearest-neighbour spin ice (blue line) is reproduced by a random walker at filling fractions $p=0.43$ and $p=0.46$ for the PC and SPC respectively. Results for filling fraction $p=0.50$, corresponding to the number of blocked spins in bSM spin ice, are shown in the right panel. A monopole moving with SM dynamics has a PSD that decays approximately as $\nu^{-2}$ at all frequencies (left panel; orange line).}
\end{figure}

In Fig.~\ref{fig:SI+PC}, the PSD of a single mobile monopole (see Methods) with bSM dynamics in ${\cal H}_{\mathrm{NN}}$ is shown together with the PSD of a random walker moving on the percolation cluster on the diamond lattice at various filling fractions. At high frequencies, the PSD of the monopole matches that of a random walker on the percolation cluster at $p_c$, and decays as $\nu^{-1.5}$. On longer time scales, the monopole moves on length scales greater than $\xi$, at which point the lattice is no longer fractal and normal diffusive behaviour is recovered. This leads to the $\nu^{-2}$ decay observed at low frequencies. From the cross-over between the two decay regimes, we again conclude that the fractal motion of the monopole occurs on time scales smaller than approximately $10^3 \ \tau_{\mathrm{fast}}$.

%
%
%
%

In the bSM version of spin ice, the ice rule constraint blocks $1/4$ of the spins (if double monopoles are energetically ruled out) and the transverse field constraint blocks $1/3$ of the remaining spins, leaving the system at an effective bond filling $p = 1/2$. While this value is larger than $p_c$ and one would thus expect a three-dimensional cluster, the blocked spins are not uncorrelated; most prominently, there is always at least one blocked spin at every site. This type of correlation has been shown to increase the percolation threshold in $d=2$ \cite{prakash1992}, and it is reasonable to expect an effective percolation threshold for bSM spin ice larger than that of uncorrelated bond correlation.

\begin{figure}
    \centering
    \includegraphics[width=1.\columnwidth]{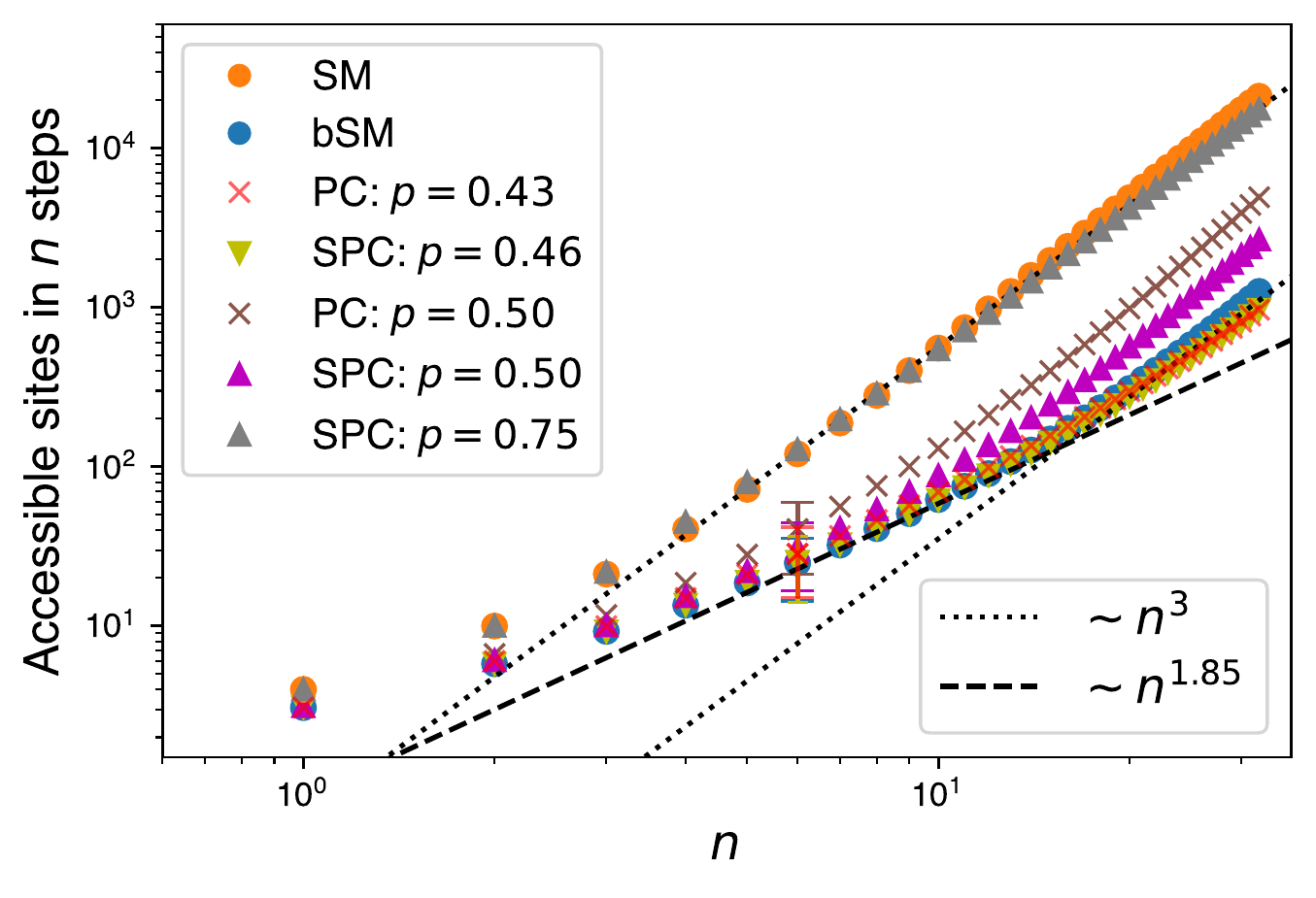}
    \caption{\label{fig:SitesReachedApp} Number of accessible sites within chemical distance $n$ of a random reference site. Results are included for a monopole on spin ice, SM/bSM (circles); for uncorrelated percolation, PC (crosses); and for percolation with at most three bonds at every vertex, SPC (up and down triangles; note that the down triangles are almost perfectly hidden by the red crosses). The grey triangles show the behaviour when every vertex has exactly $3$ bonds, distributed at random. Results for filling fraction $p=0.50$, corresponding to the number of blocked spins in bSM spin ice, are shown in brown crosses and magenta triangles. The filling fractions that give the best correspondence with bSM dynamics, $p=0.43$ for the uncorrelated percolation cluster and $p=0.46$ for the structured percolation cluster, are shown in red crosses and yellow triangles respectively.}
\end{figure}

Indeed, a walker on the random-bond percolation cluster at $p=1/2$ shows a smaller regime of fractal behaviour as compared to a monopole in bSM spin ice (see Fig.~\ref{fig:SitesReached} in the main text, and Fig.~\ref{fig:SitesReachedApp}). The bSM results are instead well-reproduced around $p=0.43$. 

%
%
%
%

Following the method described above, one can ensure that no vertex in the percolation model is connected by more than three bonds (generating structured percolation clusters, SPC). Removing $1/4$ of the bonds in this way reproduces the cluster growth of SM spin ice (see Fig.~\ref{fig:SitesReachedApp}). Reducing the filling fraction further one finds that the bSM results are now reproduced around $p=0.46$, i.e.\ closer to $0.50$. It is reasonable to assume that the transverse field distributions contribute some further correlations, and that these are responsible for the remaining discrepancy. We have not studied these correlations and their effects further, and they remain a potential topic of future research. 

It is perhaps worthy of notice how good an agreement can be found between magnetic noise from single monopole motion in bSM nearest-neighbour spin ice and an appropriately correlated bond percolation problem. Recall that the kinematic constraints in the former depend on the local spin correlations, which evolve in time; on the contrary, the bond percolation problem is entirely static (quenched disorder). However, the difference appears to be minimal within the time scales and system sizes accessed in our study.

\subsection{Monopoles and percolation clusters}
Here we also take the opportunity to expand on some of the less obvious properties of monopoles with bSM dynamics, and how they compare to the behaviour of a random walker on a percolation cluster. Firstly, one should keep in mind that the percolation structure in spin ice is only meaningful in relation to the monopoles. Furthermore, there is not a single percolation structure; every monopole has its individual set of spins it can move through and sites it can reach. As there is never a single monopole in the system, these sets are also continuously changed by the motion of other monopoles, and they may block and unblock paths for each other. In fact, this can lead to monopoles being constrained to a small number of sites, where they are either forced to remain until another monopole moves by to open up a new path for them, or until they annihilate. For ${\cal H}_{\mathrm{NN}}$ at $T=0.6$~K, we have found that at any given moment in time approximately $17$\% of the monopoles are constrained to a small number of sites ($\lesssim 60$), whereas the remaining monopoles are able to access approximately $95$\% of the sites in the system. Monopoles constrained to a small number of sites generate a small contribution to the magnetic noise, and only at relatively high frequencies. As these are also less numerous, we have chosen to focus our analysis on monopoles moving on the \quotes{percolation cluster}, i.e., the ones that can access a majority of the system sites. 

\begin{figure}
    \centering
    \includegraphics[width=1.\columnwidth]{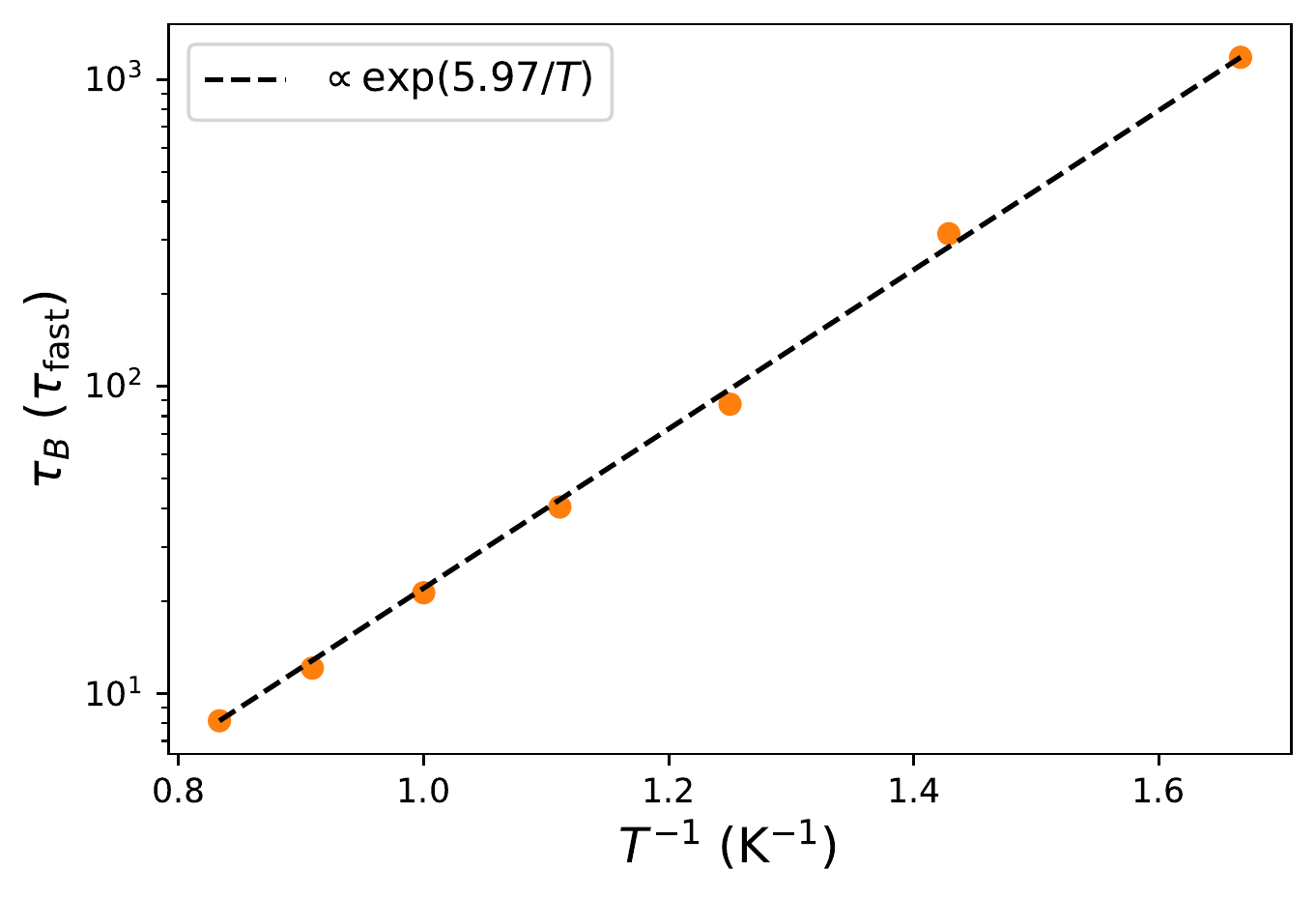}
    \caption{The characteristic time $\tau_B(T)$ on which the local transverse field configuration changes. Extracted from fitting the distribution of times $\Delta_t$ between consecutive changes of the transverse field across a spin with a function $C\exp{\left[\Delta_t/ \tau_B(T) \right]}$.}
    \label{fig:Tau_B}
\end{figure}

How fast the spin configuration around a monopole changes, i.e., the time scale on which the monopole's fractal cluster evolves, is determined by the monopole density. A lower estimate of the time scale on which the fractals evolve is given by measuring the time $\Delta_t$ between consecutive changes of the transverse field across a spin. $\Delta_t$ is distributed according to a temperature dependent exponential distribution $\sim\exp{\left(\Delta_t/ \tau_B(T) \right)}$, from which we can extract a temperature dependent characteristic time $\tau_B(T)$ (see Fig.~\ref{fig:Tau_B}). $\tau_B(T)$ is the typical time scale on which the local transverse field evolves, and we find that $\log\left(\tau_B\right) \propto A/T$, with $A\approx 6$~K ($A$ is similar to the monopole generation cost). 

%
%
%
%

The evolution of the fractal cluster destroys the memory in the system, and will set a time scale above which the monopole motion becomes diffusive. However, as this time scale grows exponentially with inverse temperature, it  becomes asymptotically unimportant when compared to the dominant effect that limits anomalous scaling: the fact that the system sits above the percolation threshold.  How memory is destroyed is explained further below, where we discuss the effect of loop updates as an example of this.


\subsection{Local one-dimensional motion}
It is interesting to note that, between blocked spins in the absence of double monopoles and spins that are prevented from flipping due to vanishing internal transverse fields, a monopole in bSM spin ice moves on average along a random one-dimensional path embedded on a three-dimensional diamond lattice. Remarkably, random walk motion $n(t)$ along such path $\vec{x}_n$ produces indeed anomalous diffusion: $\langle x^2_{n(t)} \rangle \sim \sqrt{\langle n^2(t) \rangle} \sim \sqrt{t}$ and a PSD that decays as $\nu^{-1.5}$. (We thank Shu Zhang at the Max Planck Institute for the Physics of Complex Systems for pointing out a possible analogy to random walks on random one-dimensional paths.) 
This is suggestive of a potential connection. However, the relevance of correlations (e.g., self-avoidance) in one-dimensional-like paths in monopole motion in bSM spin ice and/or on a critical percolation cluster remains to be assessed, as well as the importance of a finite density of dead ends and branching points which ought to be considered in these systems. It is an intriguing direction for future work.


\subsection{\label{app:VaryTau+Loops}
Effects of finite $\tau_{\mathrm{slow}}$ and loop updates}
\begin{figure}[h]
    \centering
    \includegraphics[width=1.\columnwidth]{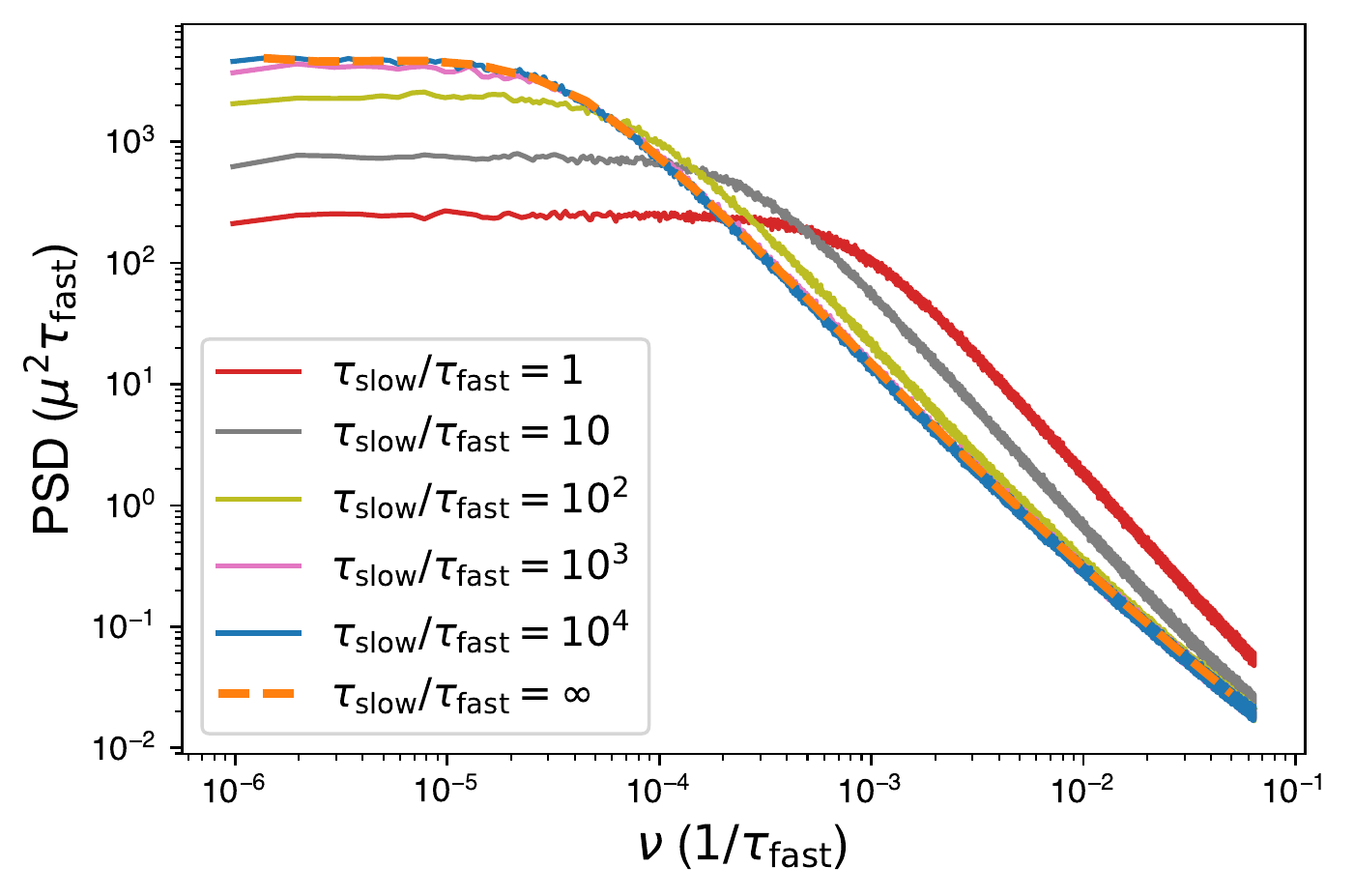}
    \caption{\label{fig:vary_tauslow}
    PSD computed for ${\cal H}_{\mathrm{NN}}$ in Monte Carlo simulations for different values of $\tau_{\mathrm{slow}}$, at $T=0.7$~K, where  $\tau_{\mathrm{slow}}/\tau_{\mathrm{fast}}=1$ is equivalent to SM dynamics. For $\tau_{\mathrm{slow}} \gtrsim 10^3 \tau_{\mathrm{fast}}$ the spins with no transverse field can be well approximated as static, as demonstrated by comparison to the dashed orange curve, which shows the PSD for the bSM where updates of the slow spins are not allowed.}
\end{figure}

In Fig.~\ref{fig:vary_tauslow}, the PSD computed with bSM dynamics is shown for different ratios $\tau_{\mathrm{slow}}/\tau_{\mathrm{fast}}$ between $1$ (equivalent to SM dynamics) and $10^4$ (approximately the ratio predicted for Dy$_2$Ti$_2$O$_7$ in Ref.~\cite{tomasello2019}, using ${\cal H}_{\mathrm{NN}}$ at temperature $T = 0.7$~K. 

The same figure also shows the PSDs obtained when updates of the slow spins are altogether forbidden, demonstrating that it is a good approximation within the time scales of interest to our study for $\tau_{\mathrm{slow}} / \tau_{\mathrm{fast}} \gtrsim 10^3$. For other spin ice compounds, the difference between the two time scales may be significantly smaller. For example, the ratio predicted for HTO is $\tau_{\mathrm{slow}} / \tau_{\mathrm{fast}} \approx 10^2$~\cite{tomasello2019}, and both time scales ought to be accounted for in the simulations. 

%
%
%
%

\begin{figure}
    \centering
    \includegraphics[width=1.\columnwidth]{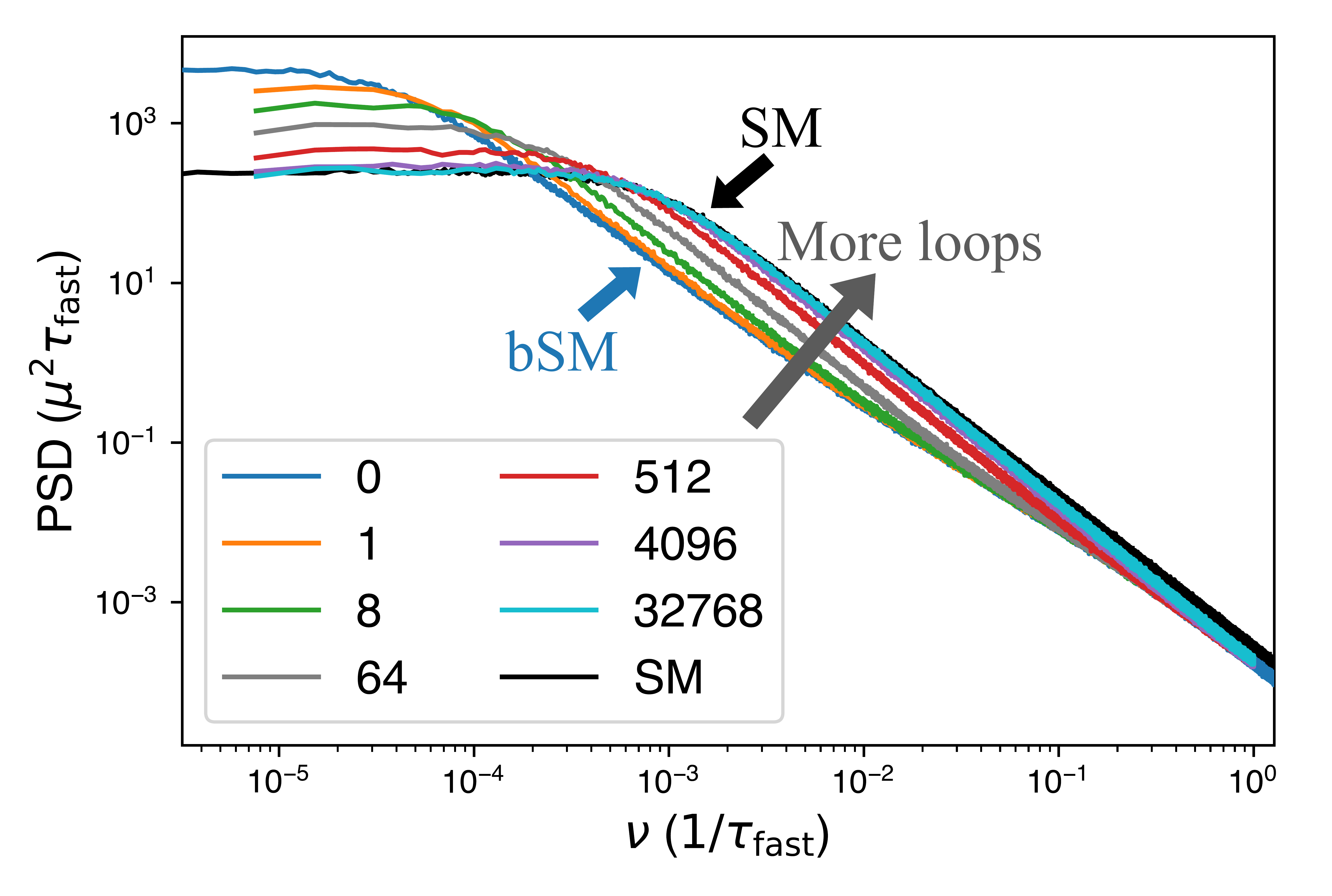}
    \caption{\label{fig:PSD-Loops}
    Applying closed loop updates in addition to single spin flip attempts in bSM spin ice removes `spin memory' from the system and approximately recovers the noise curve of SM dynamics (black curve). The coloured lines here show the PSD of bSM dynamics with different numbers of hexagonal loop updates applied per Monte Carlo sweep, varying from $0$ to $8^5$. These results were obtained in thermal equilibrium at $T=0.7$~K for ${\cal H}_{\mathrm{NN}}$, and we used a system consisting of $8192$ spins.}
\end{figure}

Evidence that the blocked spins retain memory, and that this is key to generating the fractal structures the monopoles move on, becomes apparent if we alternate bSM single spin flip dynamics with closed loop worm updates that do not change the magnetisation nor move monopoles. In this case, we notice that the PSD becomes less and less anomalous (see Fig.~\ref{fig:PSD-Loops}). In the results presented here, we only considered hexagonal loops consisting of six spins, but the general result remains unchanged if one includes larger closed loops. To approximately recover the noise of SM dynamics for a system consisting of $16\times 8^3=8192$ spins, it is sufficient to apply $4096$ hexagonal loop updates per Monte Carlo sweep. However, the noise already becomes less anomalous when one applies even a single hexagonal loop update per sweep. 
This suggests that the magnetic noise in quantum spin ice, where a significant rate of loop updates is expected, may be less anomalous -- an effect that could be taken as evidence of quantum coherent ring exchange processes in candidate quantum spin ice systems. 

%
%
%
%


\end{document}